# IL «PROTONE NEUTRO»
# OVVERO DELLA LABORIOSA ESCLUSIONE
# DEGLI ELETTRONI DAL NUCLEO

*Alberto De Gregorio**
*Dottorato di ricerca in Fisica*
*Università degli studi di Roma «La Sapienza»*

RIASSUNTO: Nel presente lavoro si discuterà l'ingresso del neutrone – o «protone neutro», come lo chiamava Ettore Majorana – nella fisica del nucleo. Si approfondiranno le circostanze in cui la scoperta del nuovo componente neutro del nucleo maturò, rilevando come la posizione assunta in principio da Irène Curie e da Frédéric Joliot nel tentare di ricondurre nell'alveo delle interpretazioni correnti i nuovi risultati da loro stessi ottenuti non fosse del tutto ingiustificata e come, per converso, la possibile esistenza del neutrone annunciata da Chadwick non bastasse a esaurire la complessità del quadro che gli esperimenti andavano mettendo in luce.

 Questione essenziale fu quella di stabilire se il neutrone fosse, come immaginato fin dal 1920, composto da un protone e un elettrone strettamente legati, o se dovesse essere considerato esso stesso particella fondamentale. Vedremo che, per oltre un anno, anche Chadwick propose per l'interpretazione del neutrone composto e sarà evidenziato il ruolo decisivo che, con Majorana ed Enrico Fermi, la scuola di fisica romana svolse nel sostenere e diffondere la concezione per cui il neutrone, così come il nucleo, non conteneva alcun elettrone.

«Protone neutro» è il nome con cui Ettore Majorana chiamò la nuova particella neutra, di massa confrontabile con quella del protone, annunciata da James Chadwick il 17 febbraio 1932.[1] Questa data, in cui il neutrone fu incluso tra le possibili particelle costituenti il nucleo atomico, è comunemente indicata come il giorno in cui l'annosa questione riguardante la presunta esistenza di elettroni nucleari trovò finalmente soluzione. Il percorso che portò a escludere l'esistenza di elettroni nel nucleo, tuttavia, ebbe una storia molto più articolata e per quasi due anni la possibilità che il neutrone risultasse dall'unione di un protone e un elettrone continuò a essere vivacemente dibattuta tra i fisici (i quali, alla scoperta dell'elettrone positivo, presero in considerazione persino l'eventualità che il protone consistesse nell'unione di un neutrone e un positrone).

Le vicende legate al neutrone – dalla sua osservazione al suo affermarsi quale particella fondamentale, non composta –, così come sono solitamente presentate, costituiscono perciò un esempio di 'linearizzazione' a posteriori della storia della fisica: obbedendo a un principio di economia, accade che lo sforzo compiuto per trarre dall'esperienza di laboratorio una formulazione generale, chiara e rigorosa di modelli e teorie passi in secondo piano, per lasciare spazio alla sola descrizione del risultato che da tale impegno si è prodotto.

---

* alberto.degregorio@roma1.infn.it

A questa estrema semplificazione contribuiscono senz'altro le numerose testimonianze rese, anche a distanza di molti anni, dagli stessi protagonisti dell'impresa scientifica. Stimolati dal notevole bagaglio 'tecnico' di cui dispongono, questi ultimi tendono a privilegiare, nel ricostruire le vicende che hanno vissuto personalmente, la sintesi raggiunta dai successivi sviluppi disciplinari e a fornire così una ricostruzione 'logica' piuttosto che storica. A maggior ragione la restante comunità scientifica, se trascura di prestare adeguata attenzione alla prospettiva sincronica delle vicende, non potrà che favorire il consolidarsi e il diffondersi di simili semplificazioni. Come un alpinista che, di ritorno dalla vetta conquistata, riferisce l'ebbrezza dell'impresa portata a compimento piuttosto che le scelte difficoltose che ha dovuto attuare nell'aprirsi una via di salita, così il fisico che ha preso parte alle pionieristiche ricerche sui neutroni tralascia facilmente di testimoniare quali accidentati percorsi abbiano permesso di raggiungere risultati di grande rilievo.

Per ottenere un più completo quadro degli eventi, rispetto a quanto emerge dalle numerose ricostruzioni diacroniche, è pertanto necessario sottoporre al riscontro della documentazione originale i resoconti dei testimoni diretti. Tale documentazione è costituita sia da materiali a stampa, come per esempio articoli, memorie e relazioni a congressi, sia da materiali d'archivio: appunti, lettere, quaderni (di laboratorio oppure contenenti sviluppi formali e calcoli relativi a determinati risultati teorici). Così come i chiodi lasciati nella roccia documentano in modo oggettivo la via scelta dallo scalatore e permettono di dare conto in maniera più completa dell'impresa da lui portata a termine, nonché di trarne insegnamenti utili a chi vorrà apprenderne le strategie, così il materiale d'epoca costituisce un necessario complemento alle testimonianze dei fisici che si sono impegnati nelle ricerche sui neutroni e consente di apprezzare più pienamente il significato dei risultati da loro conseguiti in questo campo.

Nel presente lavoro, avvalendoci del materiale a stampa, ricostruiremo alcune vicende legate alla scoperta del neutrone e al percorso verso la sua accettazione come particella fondamentale, priva di elettroni. Accennerò in apertura al fatto che l'iniziale interpretazione in termini di radiazione γ fornita da Frédéric Joliot e da sua moglie Irène Curie per la radiazione penetrante emessa dal berillio trova una naturale collocazione nelle conoscenze dell'epoca, rispetto alle quali presenta riscontri anche quantitativi. Sarà poi approfondito il fatto che dopo gli esperimenti di Chadwick ci si interrogò ancora a lungo se il neutrone consistesse nella stretta unione di un protone e un elettrone, e sarà evidenziato il ruolo fondamentale svolto prima da Majorana e poi da Enrico Fermi nel promuovere l'interpretazione per cui il neutrone andava considerato a tutti gli effetti un «protone neutro», non contenente – come il nucleo stesso – alcun elettrone.

1. LA RADIAZIONE PENETRANTE

A scoprire l'esistenza della radiazione penetrante di origine nucleare furono nel 1930 Wilhelm Bothe e Herbert Becker,[2] impegnati ad approfondire il fatto che, come osservato nel 1928 dallo stesso Bothe insieme a Hans Fränz,[3] il boro irradiato con particelle α emetteva nuclei di idrogeno (ossia protoni) che potevano essere classificati secondo due possibili valori della loro velocità. Bothe e Fränz avevano ipotizzato che in tale processo, oltre a nuclei di idrogeno, il boro emettesse radiazione γ e che in questo modo fosse garantito il bilancio energetico (considerazioni sui difetti di massa[4] escludevano che protoni con velocità differenti provenissero da due diversi isotopi del boro, di numero di massa 10 e 11). Bothe e Becker intrapresero (mediante contatori a punta) lo studio di tale radiazione non soltanto nel boro, ma anche in altri elementi leggeri sottoposti all'azione delle α, tra cui litio, berillio, fluoro, magnesio e alluminio. Con stupore constatarono che radiazione era emessa anche dal berillio, l'unico tra i precedenti elementi da cui non risultava essere espulso alcun protone. L'assorbimento della radiazione eccitata nel berillio dalle particelle α risultava minore rispetto anche ai più penetranti tra i raggi emessi dalle sostanze radioattive naturali; da qui il nome, appunto, di «radiazione penetrante del berillio». Era necessario specificarne la provenienza, per distinguerla dalla radiazione penetrante *tout court* di cui si era già parlato con riferimento ai raggi cosmici.

Nell'ottobre del 1931 si tenne a Roma un importante congresso di fisica nucleare, al quale parteciparono tra gli altri N. Bohr, W.W. Bothe, R.H. Fowler, G. Gamow e A. Sommerfeld. Fermi fu segretario generale del convegno, ai cui lavori assistette anche Ettore Majorana (come documenta unicamente una foto che lo ritrae insieme ai colleghi). Bruno Rossi discusse *Il problema della radiazione penetrante*[5] di origine cosmica e precisò che, pur continuando in questo caso a parlare di radiazione *corpuscolare* penetrante, avrebbe mantenuto la riserva (avanzata in precedenza anche da Bothe e Kolhörster[6]) che si trattasse eventualmente di radiazione elettromagnetica, di energia non inferiore al miliardo di elettronvolt. In tale ipotesi, tuttavia, si sarebbero dovute attribuire a raggi così altamente energetici proprietà diverse rispetto a quelle estrapolate da leggi la cui validità sperimentale era stata verificata a energie ordinarie. Da una parte, infatti, i risultati di alcune misurazioni portavano a concludere che, se la radiazione cosmica primaria fosse stata di natura elettromagnetica, ciascun fotone avrebbe percorso nell'aria in media alcuni centimetri solamente. Dall'altra parte la formula che Oskar Klein e Yoshio Nishina avevano ricavato teoricamente per la sezione d'urto di radiazione elettromagnetica contro elettroni[7] – formula le cui previsioni erano in buon accordo con i risultati sperimentali fino a energie di alcune centinaia di migliaia di elettronvolt – prevedeva per il percorso di fotoni di così straordinaria energia un valore di diversi chilometri, evidenziando pertanto una differenza di cinque ordini di grandezza rispetto al dato osservativo.

Riguardo all'altra radiazione penetrante – quella emessa dal berillio irradiato con le α –, Bothe[8] annunciò che alcuni esperimenti intrapresi con il metodo delle coincidenze suggerivano che essa avesse natura elettromagnetica e non corpuscolare.

Alcune settimane dopo il congresso di Roma i Joliot-Curie,[9] ricorrendo a sorgenti di polonio molto più intense di quelle usate da Bothe e Becker nel 1930, studiarono mediante una camera di ionizzazione la radiazione penetrante, emessa, oltre che dal berillio, anche dal boro e dal litio: l'assorbimento per unità di massa della radiazione del berillio risultava essere nel piombo di entità minore rispetto a quanto stabilito dai colleghi di Berlino.[10] I due fisici francesi inoltre, intenti a verificare se avvenissero trasmutazioni nucleari con emissione di particelle α oppure di protoni, osservarono che la radiazione penetrante del berillio e del boro espelleva protoni da schermi composti di sostanze idrogenate come paraffina, acqua o cellofan, collocati lungo il percorso.[11]

I Joliot-Curie, nella seduta del 22 febbraio 1932 dell'Académie des Sciences, presentarono un altro lavoro, in cui riferirono di aver osservato in camera di Wilson alcune traiettorie prodotte da nuclei di idrogeno di rinculo, oltre a tracce di elettroni di elevata energia. I due fisici, ricorrendo nuovamente alla camera di ionizzazione, avevano constatato che la radiazione emessa dal berillio sottoposto all'azione delle particelle α del polonio è fortemente assorbita dal carbonio e, in una camera contenente elio, provoca un aumento della corrente misurata: da tale risultato concludono che «le phénomène de projection des noyaux d'atomes par les rayons γ de grande énergie est probablement un phénomène général».[12]

Per quanto riguarda questi esperimenti, i Joliot-Curie devono affrontare gravi difficoltà interpretative: l'energia dei nuclei di idrogeno di rinculo è tale da implicare che, se la radiazione penetrante del berillio e del boro fosse costituita di fotoni, questi dovrebbero avere energie rispettivamente di 48 MeV e 31 MeV, molto più di qualsiasi radiazione γ nota di origine nucleare; in contrasto con queste conclusioni, studi di assorbimento portano a concludere che la radiazione penetrante debba invece, se di natura elettromagnetica, essere costituita di fotoni con energie tra i 15 e i 20 MeV. In aggiunta, la formula di Klein-Nishina prevede per fotoni di tali energie che l'ipotizzato effetto Compton su protoni abbia una sezione d'urto circa $10^5$ volte più piccola che non l'effetto Compton su elettroni,[13] ma i Joliot-Curie deducono da risultati di laboratorio che l'interazione con i protoni è addirittura più frequente di quella con gli elettroni. Lo scarto da loro riscontrato rispetto alla previsione teorica ammonta perciò a circa cinque ordini di grandezza; ossia è di entità analoga a quello evidenziato da Rossi per il percorso medio della radiazione cosmica primaria, proprio nell'ipotesi che quest'ultima sia di natura elettromagnetica.[14]

Gli esperimenti, nel caso in cui la radiazione penetrante del berillio fosse stata interpretata in termini di fotoni, conducevano dunque a conclusioni non solamente contrastanti tra loro, ma anche non conciliabili con la teoria di Klein e Nishina. I Joliot-Curie, assumendo un atteggiamento propositivo, ne dedussero la possibile manifestazione di «un noveau mode d'interaction du rayonnement et de la matiére».[15] Tale conclusione, a prima vista fin troppo audace, aveva invece valide basi osservative, di solito non poste sufficientemente in evidenza: in primo luogo Bothe,[16] in un intervento che Irène Curie citò espressamente a dicembre,[17] durante il congresso di Roma aveva messo in risalto la natura elettromagnetica della radiazione del berillio, emessa a suo avviso in conseguenza della «Kernsynthese» $Be^9 + He^4 = C^{13}$ (di lì ad alcuni mesi Franco Rasetti[18] avrebbe verificato che la radiazione del berillio conteneva, effettivamente, non soltanto neutroni, ma anche raggi $\gamma$). In secondo luogo, proprio durante il congresso romano era stato puntualizzato che, se si fosse accettata l'ipotesi elettromagnetica per i raggi cosmici primari, si sarebbe dovuto ammettere per la radiazione un nuovo modo di interagire con la materia: negli articoli che seguirono, i Joliot-Curie, pur non dichiarandolo esplicitamente, non fecero altro che trarne conclusioni speculari riguardo alla radiazione penetrante del berillio. Non va peraltro dimenticata la coincidenza per cui, nell'ipotesi di radiazione elettromagnetica di elevatissima energia, la sezione d'urto misurata si discostava da quella prevista dalla formula di Klein-Nishina – verificata sperimentalmente soltanto a energie assai inferiori – in analoga proporzione per la radiazione penetrante del berillio e per quella di origine cosmica.

C'è da segnalare per completezza che i Joliot-Curie, nell'optare per l'interpretazione elettromagnetica, citarono esplicitamente non soltanto i risultati annunciati a Roma da Bothe e da lui ottenuti con la tecnica delle coincidenze, ma anche le numerose tracce di elettroni di alta energia che essi stessi avevano osservato in camera di Wilson: la spiegazione più semplice era che tali elettroni fossero stati prodotti mediante effetto Compton da fotoni provenienti dal berillio (come in buona parte effettivamente è, a posteriori).

La soluzione alle contraddizioni sorte attorno alla radiazione penetrante del berillio si prospettò in maniera inizialmente molto incerta, come emerge per esempio da un lavoro di H.C. Webster, comunicato da Chadwick alla Royal Society nel gennaio del 1932: studi di assorbimento, condotti con l'ausilio di una camera a ionizzazione, secondo Webster sono sufficientemente in accordo con le previsioni teoriche da accreditare l'ipotesi che la ionizzazione osservata sia prodotta da radiazione elettromagnetica, e non «by high-speed corpuscles consisting, *e.g.*, of a proton and an electron in very close combination»;[19] egli tiene anche presente che alcune fotografie in camera di Wilson non mostrano tracce che siano riconducibili, direttamente o indirettamente, a simili corpuscoli neutri, noti già allora con il nome di neutroni. E tuttavia, a Webster non sfugge il fatto che la differenza di energia che sembra esservi tra la radiazione emessa dal berillio nel verso delle particelle $\alpha$ incidenti e in quello opposto è di entità tale da rappresentare un serio ostacolo all'interpretazione elettromagnetica.[20]

Il 17 febbraio 1932, Chadwick propone la chiave interpretativa che risolve le difficoltà e le contraddizioni che sono sorte dal considerare la radiazione penetrante costituita di sola radiazione elettromagnetica e, in una celebre lettera a Nature dal titolo *Possible Existence of a Neutron*,[21] annuncia una serie di risultati che suggeriscono l'esistenza di «particles of mass 1 and charge 0, or neutrons»: Chadwick ha verificato con un «valve counter» (ossia con un sistema di rivelazione costituito da una piccola camera di ionizzazione collegata a un amplificatore) che la radiazione penetrante espelle particelle ionizzanti non soltanto dall'idrogeno, ma anche da elio, litio, berillio, carbonio, aria e argon. Un'assai semplice analisi quantitativa esclude che questo specifico fenomeno possa essere provocato da radiazione elettromagnetica.

Il berillio irradiato con particelle α emette dunque neutroni anziché protoni; il boro sia protoni sia neutroni. In entrambi i casi è effettivamente presente anche radiazione di natura elettromagnetica, dotata di energia superiore a quella dei γ emessi dai radioelementi, ma comunque inferiore rispetto alle diverse decine di MeV ipotizzati inizialmente.

Già da diversi anni era stata formulata l'ipotesi che nel nucleo potessero esservi corpuscoli neutri; l'esempio più noto è quello – già citato – di una stretta combinazione di un protone e un elettrone, ma non mancavano strutture più complesse. In particolare Ernest Rutherford aveva proposto nell'agosto del 1927, e discusso al congresso per le celebrazioni voltiane svoltosi il successivo settembre a Como, un modello basato sulla presenza di 'satelliti' neutri nel nucleo atomico.[22] Del modello di Rutherford, che descriveremo nel prossimo paragrafo, si occupò nel 1928 Giovanni Gentile jr, figlio del celebre filosofo.

## 2. LA TEORIA DEI SATELLITI DI RUTHERFORD

Il problema che Rutherford si propose di risolvere nel 1927 era quello per cui i dati sperimentali portavano a risultati talvolta contraddittori riguardo alla costituzione del nucleo. Negli esperimenti di diffusione coulombiana, per esempio, l'accordo tra le previsioni teoriche basate sulla legge dell'inverso del quadrato della distanza e i risultati sperimentali ottenuti irradiando l'uranio con particelle α veloci (come quelle di 7,7 MeV emesse dal RaC′ [$Po^{214}$]) dimostrava che i nuclei dell'elemento numero 92 non avrebbero potuto avere dimensioni superiori a $3,2 \cdot 10^{-12}$ cm; allo stesso tempo, il valore dell'energia posseduta dalle particelle α emesse a loro volta dall'uranio (minore rispetto all'energia delle α del RaC′) indicava che la struttura nucleare dell'elemento 92 dovesse estendersi fino a distanze di almeno $6 \cdot 10^{-12}$ cm.

Il modello proposto da Rutherford nel 1927 prevede che nel nucleo orbitino alcuni «satelliti» neutri, costituiti dall'unione di una particella α e di due elettroni; tali satelliti rimangono legati al nucleo carico per effetto della polarizzazione che quest'ultimo induce su di essi e della conseguente forza attrattiva (che, nell'approssimazione che $r$ sia grande rispetto alle dimensioni lineari del satellite, ha andamento $1/r^5$). Sono le profonde deformazioni che il nucleo carico induce nella distribuzione di carica posseduta dalla particella α a giustificare il fatto che due elettroni possano trovarsi legati a quest'ultima molto più saldamente nel satellite che non nell'atomo di elio. Il sistema così descritto è meccanicamente instabile e il modello prevede (nei limiti di validità della precedente approssimazione) orbite quantizzate poste a distanze via via *decrescenti* dal nucleo carico – ossia l'opposto di ciò che accade per gli stati elettronici di un atomo –: l'energia cinetica lungo un'orbita $n$ è proporzionale a $n^4$ e il raggio a $1/n$. In occasione di una disintegrazione radioattiva la particella α è espulsa con un'energia finale pari alla metà dell'energia cinetica posseduta inizialmente e il nucleo superstite assorbe i due elettroni a cui essa era legata.

Il modello di Rutherford portava a previsioni quantitative in buon accordo con le energie delle particelle α delle principali serie radioattive e suggeriva una spiegazione anche per i raggi γ, la cui emissione sarebbe avvenuta a causa della transizione di un satellite da uno stato quantizzato a un altro.

Pochi mesi dopo la pubblicazione di questo articolo, il 5 febbraio 1928, Corbino presenta ai Lincei una nota con la quale il giovane Gentile[23] solleva seri dubbi sul modello costruito da Rutherford. Appare innanzitutto discutibile che il fisico britannico faccia ricorso a numeri quantici molto elevati (tipicamente $n \sim 25$), per di più sia interi sia seminteri. Ma soprattutto, secondo i calcoli di Gentile la vita media degli stati quantizzati sarebbe decine di ordini di grandezza inferiore rispetto a quella osservata: per esempio, mediante calcoli di elettrodinamica classica, si ottengono per lo ionio ($Th^{230}$) $0,7 \cdot 10^{-15}$ secondi, a fronte di una vita media misurata di migliaia di anni; e d'altra parte «il calcolo delle vite medie con la elettrodinamica classica ha dato sempre valori compatibili con la realtà sperimentale».[24]

Amaldi afferma che quando, sul finire del 1931, i Joliot-Curie proposero che i protoni prodotti dalla radiazione penetrante del berillio indicassero che quest'ultima era costituita di fotoni di elevata energia, Majorana commentò: «Non hanno capito niente; probabilmente si tratta di protoni di rinculo prodotti da una particella neutra pesante»;[25] anche Emilio Segrè riferisce che Majorana «immediately understood that there was what he called "a neutral proton"».[26] Amaldi riconduce[27] la felice intuizione di Majorana al fatto che Gentile, amico di Majorana, si era occupato nel 1928 della teoria dei satelliti neutri.

Majorana conosceva senz'altro il lavoro di Rutherford del 1927, non solamente per l'analisi che ne aveva fatto Gentile, ma anche per essersi occupato personalmente nella propria tesi di laurea (discussa nel 1929) del problema delle differenti dimensioni nucleari dedotte dagli esperimenti di diffusione e dai fenomeni radioattivi. Rutherford stesso, però, pur avendo proposto l'esistenza di corpuscoli nucleari neutri, nel 1927 si era preoccupato di mettere in debita luce che «it is an essential part of the present theory that the satellite must lose its two electrons before its complete escape from the nucleus»;[28] in aggiunta, «it seems likely that neutral satellites are important constituents of all elements of atomic number greater than 30, but probably cannot exist as constituent of the very light elements».[29]

È per queste due precisazioni dello stesso Rutherford – i satelliti evolvono verso l'espulsione di particelle cariche e sono ingredienti strutturali esclusivamente per nuclei pesanti – che appare quanto meno improbabile che possa essere stato il suo articolo del 1927 a suggerire a Majorana che i risultati ottenuti a cavallo tra il 1931 e il 1932 dai Joliot-Curie sul berillio andassero interpretati come emissione di una particella neutra da parte di un nucleo leggero. Influenza, se vi fu, potrebbe essere stata di natura indiretta, in quanto nel lavoro del 1927 era citato esplicitamente il neutrone, la particella della quale Rutherford aveva discusso l'esistenza ben sette anni prima, prevedendone però caratteristiche molto diverse da quelle dei satelliti neutri.

Di neutrone si era andati parlando in realtà da tempo – addirittura dall'Ottocento –, ma a questo termine erano stati attribuiti i significati più disparati (dall'unione di cariche positive e negative costituenti l'etere a quella di elettroni e positroni,[30] mentre Wolfgang Pauli avrebbe dato inizialmente il nome di «neutrone» alla particella deputata a garantire la conservazione dell'energia nel decadimento β). La possibilità di una stretta combinazione tra protone ed elettrone era stata trattata dal chimico statunitense William D. Harkins, che a metà degli anni Dieci aveva pubblicato insieme a Ernest D. Wilson alcuni lavori sulla struttura dei nuclei atomici. Il 12 aprile del 1920 il Journal of the American Chemical Society ricevette un suo articolo[31] che sintetizzava le teorie esposte in precedenza. È molto interessante constatare come Harkins, nel descrivere la struttura dei nuclei, ricorra a diversi gruppi di particelle: considera gruppi μ formati da due protoni e due elettroni $(\eta_2^+\beta_2^-)^0$, o la particella ν di massa 3, contenente tre protoni e due elettroni $(\eta_3^+\beta_2^-)^+$. Nel nucleo svolgerebbe un ruolo molto importante anche il cosiddetto «helio group», una particella α neutralizzata elettricamente mediante l'unione di due «cementing electrons»: $(\eta_4^+\beta_4^-)^0$; è evidente la sua somiglianza con il satellite neutro discusso nel 1927 da Rutherford. Inoltre, secondo il chimico, «it is not improbable that some isotopic atoms are formed by the addition of the group $(\eta^+\beta^-)^0$».[32] Harkins non attribuisce alcun nome a questo ulteriore gruppo, che però come è evidente altro non è che il neutrone di cui parlerà Rutherford alcune settimane più tardi.[33]

Il 3 giugno del 1920 Rutherford tenne la Bakerian Lecture dal titolo *Nuclear Constitution of Atoms*,[34] nella quale il nucleo è descritto come costituito di protoni ed elettroni (eventualmente raggruppati a formare particelle α); tali elettroni, a causa delle intense forze a cui sono sottoposti, risultano «deformed» al punto da rimanere legati nel nucleo. Il fisico britannico ripercorre gli esperimenti che nel 1919 lo hanno portato a scoprire che il nucleo di azoto si trasforma in uno di ossigeno e uno di idrogeno, secondo la reazione:

$$N_7^{14} + He_2^4 \rightarrow O_8^{17} + H$$

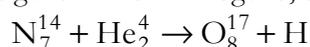

Il protone è così annoverato tra i costituenti nucleari, in ossequio al principio per cui particelle espulse dal nucleo preesistono nel nucleo stesso (principio la cui validità verrà meno soltanto con l'affermarsi della teoria del decadimento β pubblicata da Fermi a cavallo tra il 1933 e il 1934). L'ossigeno e l'azoto, inoltre, sembrano emettere per azione delle α alcune particelle costituite da tre protoni e un elettrone; quindi, aggiunge Rutherford, perché non ipotizzare che «one electron can also bind two H nuclei and possibly also one H nucleus»? In quest'ultimo caso ci si troverebbe al cospetto di un «atomo» di massa uno e carica nulla, con proprietà del tutto peculiari:

> Under some conditions […] it may be possible for an electron to combine much more closely with the H nucleus, forming a kind of neutral doublet. Such an atom would have very novel properties. Its external field would be practically zero, except very close to the nucleus, and in consequence it should be able to move freely through matter.[35]

A differenza dei satelliti nucleari del 1927, il «doppietto neutro» del 1920, costituito da un protone e un elettrone, secondo lo stesso Rutherford sarebbe in grado di abbandonare il nucleo senza scindersi nelle proprie parti cariche costituenti e potrebbe attraversare pressoché indisturbato la materia. Nella Bakerian Lecture Rutherford, come già Harkins, non affronta la questione del nome; il termine neutrone riferito a questo doppietto neutro è citato invece, ad esempio, nella riunione della British Association del 25 agosto 1920.[36]

Negli anni Venti Chadwick affrontò, a più riprese ma senza successo, la ricerca sperimentale del neutrone di Harkins e Rutherford. Nel 1928, come già ricordato, si scoprirono i presunti raggi γ emessi dal boro irradiato con le α; quindi si giunse alla scoperta della radiazione penetrante, agli esperimenti dei Joliot-Curie e a quelli decisivi di Chadwick, che nel 1932 poté finalmente annunciare la *Possible Existence of a Neutron*. Il neutrone di Harkins e Rutherford, osservato da Chadwick, non era concepito però come particella fondamentale, ma era l'unione di un protone e di un elettrone. Non un 'protone neutro', quindi, ma una coppia di particelle che si neutralizzavano elettricamente a vicenda, o, se vogliamo, un 'protone neutralizzato'.

Secondo quanto riferisce Amaldi, «quello che è certo è che prima della Pasqua [del 1932 Majorana] aveva cercato di fare la teoria dei nuclei leggeri ammettendo che i protoni e i neutroni (o protoni neutri come lui diceva allora) ne fossero i soli costituenti e che i primi interagissero con i secondi con forze di scambio».[37] La questione terminologica tra il neutrone di Harkins e Rutherford (o 'protone neutralizzato') e il 'protone neutro' di Majorana ben riassume il dilemma sulla natura della nuova particella e sulla presenza di elettroni nel nucleo.

3. ELETTRONI NEL NUCLEO
Le difficoltà che scaturivano dal considerare il neutrone come una stretta combinazione di protone ed elettrone riflettevano quelle inerenti al modello di nucleo al cui interno i componenti carichi positivamente (particelle α, protoni) coesistevano con alcuni elettroni negativi.

La concezione secondo la quale il nucleo conteneva elettroni era andata affermandosi, grazie in particolare al contributo di Marie Curie e di Niels Bohr, nei primi anni Dieci del Novecento, in relazione ai fenomeni radioattivi di tipo β:[38] costituiva la necessaria sintesi tra il principio per cui gli elettroni espulsi dovevano preesistere in qualche regione dell'atomo e, dall'altro lato, il dato osservativo che dall'esterno non era possibile intervenire in alcun modo sugli elettroni emessi nel decadimento β (mentre era possibile agire sugli elettroni orbitali).

Nel quadro della meccanica quantistica, però, ammettere l'esistenza di elettroni nel nucleo implicava, come è noto, gravi difficoltà: esse andavano dal non poter dare conto del fatto che particelle tanto leggere rimanessero confinate in regioni ristrette come quelle nucleari (paradosso di Klein) alla contraddizione che si era manifestata, con la scoperta e le prime applicazioni dell'effetto Raman, tra i risultati sperimentali e quelli teorici relativamente allo spin dei nuclei di azoto e successivamente anche alla loro statistica.[39]

Walter Heitler e Gerhard Herzberg sintetizzarono: «Scheint es, als ob das Elektron im Kern mit seinem Spin auch sein Mitbestimmungsrecht an der Statistik des Kerns […] verliert».[40] Tre anni più tardi, con l'annuncio della possibile esistenza del neutrone, la situazione non sarebbe cambiata nella sostanza: elettroni avrebbero continuato ad essere necessari nel nucleo né più né meno di quanto non lo fossero stati fino ad allora e il neutrone di Harkins, Rutherford e Chadwick, costituito di un protone e un elettrone, avrebbe sollevato le medesime difficoltà che erano state poste da nuclei contenenti elettroni 'liberi'. Tuttavia, al pari di Heitler e Herzberg, nel 1930 V. Ambarzumian e D. Iwanenko sposarono l'idea che nel nucleo gli elettroni perdessero «leur individualité [et] ces considerations nous ont amenés à essayer de construire une théorie des rayons β».[41] Nel 1932 Iwanenko tornò in maniera più estesa sull'argomento e concluse: «Nous ne considérons pas le neutron comme constitué d'un électron et d'un proton mais comme une *particule élémentaire*».[42]

In altre parole, l'osservazione del neutrone, sebbene lasciasse ancora aperta la questione degli elettroni nucleari, tuttavia ridestò – o suscitò – in alcuni l'attenzione per un neutrone inteso come particella 'fondamentale', o 'elementare'; con tutte le conseguenze che ciò comportava per il nucleo.

Della questione impostasi all'attenzione generale nel febbraio del 1932 riguardo al neutrone di Chadwick, se considerarlo un doppietto alla maniera di Harkins e Rutherford o come un'unica particella, è rimasta ampia documentazione, della quale riporteremo alcuni brani a titolo esemplificativo.

In un articolo apparso sui Proceedings of the Royal Society il 10 maggio 1932, Chadwick stesso si espresse in questi termini (non del tutto inequivoci):

> We may suppose it to consist of a proton and an electron in close combination, the «neutron» discussed by Rutherford in his Bakerian Lecture of 1920 […]. We may then proceed to build up nuclei out of α-particles, neutrons and protons, and we are able to avoid the presence of uncombined electrons in a nucleus. This has certain advantages for, as is well known, the electrons in a nucleus have lost some of the properties which they have outside, e.g., their spin and magnetic moment […]. It has so far been assumed that the neutron is a complex particle consisting of a proton and an electron. This is the simplest assumption […]. It is, of course, possible to suppose that the neutron may be an elementary particle. This view has little to recommend at present.[43]

Un anno e mezzo più tardi, tra i partecipanti al Congresso Solvay prevaleva in linea di massima l'orientamento a considerare il neutrone come una unica particella, ma la questione rimaneva completamente aperta, tanto che Bohr arrivò a esprimersi in questi termini: «A mon avis, le sens qu'on doit attacher à la distinction entre particules élémentaires et particules complexes ne peut pas être indiqué sans ambiguïté».[44] In modo analogo si discusse della possibilità che il neutrone esistesse in stati di massa differente[45] o che il protone risultasse a sua volta dall'unione di un elettrone positivo e di un neutrone: Carl Anderson, in seguito alla scoperta del positrone, a febbraio aveva avanzato l'ipotesi della complessità del protone, ripresa dai Joliot-Curie a giugno. Lo stesso Fermi aveva discusso tale modello di protone con Gleb Wataghin, prima che il 23 aprile 1933 questi presentasse ai Lincei una nota sulla teoria del nucleo.[46] Al Congresso Solvay, Perrin sintetizzò il significato del protone composto, affermando che «il y a donc sans doute une symétrie complète au point de vue de la complexité entre le neutron et le proton».[47]

Per portare un altro esempio importante, è noto che anche Heisenberg, quando nel giugno del 1932 inviò alla Zeitschrift für Physik il primo dei tre articoli in cui esponeva la teoria del nucleo basata su forze di scambio, non assunse una posizione netta sulla questione degli elettroni nucleari. Possiamo chiarire meglio questo punto riproponendo alcuni passaggi: da una parte il fisico tedesco osserva che è possibile *ipotizzare* che i nuclei siano composti da protoni e neutroni ma non contengano alcun elettrone; pochi giorni dopo spiegherà a Bohr: «Die Grundidee ist: alle prinzipiellen Schwierigkeiten auf das Neutron abzuschieben und im Kern Quantenmechanik zu treiben».[48] Se si escludono gli elettroni dal nucleo, scrive quindi nel primo articolo, «die fundamentalen Schwierigkeiten, denen man in der Theorie des β-Zerfalls und der Stickstoffkernstatistik begegnet, lassen sich nämlich dann reduzierene auf die Frage, in welcher Weise ein Neutron in Proton und Elektron zerfallen kann und welcher

Statistik es genügt».[49] D'altra parte, Heisenberg poco oltre sembra tornare sui suoi passi e suggerisce un parallelo tra la forza di scambio che tiene legati i due protoni nello ione molecolare di idrogeno $H_2^+$ e quella che si esercita tra un protone e un neutrone nel nucleo: la quantità che esprime in termini matematici il legame di scambio, infatti, «kann man wieder durch das Bild der Elektronen, die keinen Spin haben und den Regeln der Bosestatistik folgen, anschaulich machen».[50] In conclusione di articolo egli aggiunge che l'espressione da lui scritta per l'energia di legame «nur gelten kann, wenn die Bewegung der Protonen langsam relativ zur Bewegung des Elektrons im Neutron erfolgt».[51] Questo approccio riemerge il 18 luglio, quando, nel discutere la diffusione di radiazione γ, Heisenberg scriverà a Bohr: «wird das einzelne Neutron, d.h. die negative Ladung in ihm, streuen».[52] Anche nel suo secondo articolo[53] prenderà in considerazione la combinazione di elettrone e protone, e in particolare la loro energia di legame. D'altronde, Eugene Wigner, nel presentare nel dicembre del 1932 la propria teoria delle forze nucleari, dimostra di aver colto chiaramente che Heisenberg si rifà al modello di neutrone composto: nello stilare una casistica delle varie teorie esistenti riguardo alle particelle elementari, il fisico ungherese cita quella in cui «the only elementary particles are the proton and the electron. This point of view has been emphasized by Heisenberg and treated by him in a series of papers».[54]

### 4. Il «protone neutro»

Come abbiamo già ricordato, Amaldi afferma che prima della Pasqua del 1932 Majorana aveva elaborato una teoria del nucleo basata su forze di scambio. Fermi, di ritorno da un convegno tenutosi a Parigi nell'estate del 1932, lesse sulle pagine della Zeitschrift für Physik la teoria di Heisenberg e probabilmente riscontrò una notevole somiglianza con ciò che Majorana gli aveva esposto privatamente alcuni mesi prima:

> [Majorana] aveva parlato di questo abbozzo di teoria [sulle forze di scambio] agli amici dell'Istituto e Fermi, che ne aveva subito riconosciuto l'interesse, gli aveva consigliato di pubblicare al più presto i suoi risultati, anche se parziali. Ma Ettore non ne volle sapere perché giudicava il suo lavoro incompleto.[55]

Prima di recarsi a Parigi, Fermi aveva chiesto a Majorana il permesso di accennare alle sue idee sulle forze nucleari, ma aveva ricevuto un nuovo, fermo rifiuto. Di ritorno dalla conferenza, «Fermi si adoperò nuovamente perché Majorana pubblicasse qualche cosa, ma ogni suo sforzo e ogni sforzo di noi suoi amici e colleghi fu vano», ricorda ancora Amaldi.[56]

A quanto pare, Fermi riuscì però a convincere il giovane fisico a recarsi a Lipsia e Copenaghen. Fu nella città tedesca che Majorana conobbe Heisenberg, con il quale strinse rapporti di amicizia e collaborazione segnati da una profonda stima reciproca. Il 18 febbraio 1933 per esempio annuncia al padre: «Ho scritto un articolo sulla struttura dei nuclei che a Heisenberg è piaciuto molto benché contenesse alcune correzioni a una sua teoria».[57] Il 22 febbraio alla madre: «Nell'ultimo "colloquio", riunione settimanale a cui partecipano un centinaio tra fisici, matematici, chimici, etc., Heisenberg ha parlato della teoria dei nuclei e mi ha fatto molta réclame a proposito di un lavoro che ho scritto qui. Siamo diventati abbastanza amici in seguito a molte discussioni scientifiche e ad alcune partite a scacchi».[58] Il 28: «Caro papà, […] mi fermerò probabilmente a Lipsia ancora due o tre giorni perché devo chiacchierare con Heisenberg. La sua compagnia è insostituibile e desidero approfittare finché egli rimane qui».[59]

A Lipsia, dunque, Majorana si era finalmente convinto a dare alle stampe la propria teoria dei nuclei. In apertura dell'articolo, il giovane fisico italiano afferma: «La scoperta del neutrone, cioè di una particella elementare pesante e senza carica elettrica, ha offerto la possibilità di edificare una teoria della struttura nucleare che, senza risolvere le difficoltà connesse con lo spettro continuo dei raggi β, permette tuttavia di utilizzare largamente i concetti della meccanica quantistica in un campo che sembrava loro estraneo».[60] In poche righe, oltre ad affermare in maniera forte e inequivocabile che il neutrone è elementare – un protone neutro a tutti gli

effetti –, Majorana sintetizza il significato più profondo che risiede nella teoria delle forze di scambio: garantire alla meccanica quantistica un più ampio accesso al nucleo.[61]

In cosa consistevano dunque le «correzioni» alla teoria di Heisenberg di cui Majorana scrive al padre il 18 febbraio? Detto che Majorana esclude categoricamente la presenza di elettroni, egli innanzitutto semplifica il formalismo introdotto da Heisenberg; quindi modifica la natura dello scambio: il fisico tedesco aveva inteso che lo scambio riguardasse sia le posizioni sia gli spin di neutrone e protone, scambio che, come egli stesso avrebbe affermato nell'ottobre 1933 al Congresso Solvay, equivaleva a quello della sola carica.[62] Ricordiamo allora che nel suo primo articolo Heisenberg aveva immaginato «lo scambio di elettroni senza spin, che obbediscono alla statistica di Bose»: considerare lo scambio della sola carica tra protone e neutrone – o della posizione e dello spin, come egli aveva formalizzato inizialmente il processo – riflette esattamente la concezione per cui due protoni si scambiano un elettrone privo di spin. In altre parole, c'è una corrispondenza stretta tra il fatto che Heisenberg invochi lo scambio di posizione e spin e il fatto che egli si rappresenti mentalmente il nucleo come contenente elettroni privi di spin.

Majorana introduce uno scambio delle sole coordinate cartesiane. Questo nuovo genere di scambio (che come osserverà Heisenberg equivale allo scambio simultaneo di carica e spin) è il più naturale se si pensa invece a protoni e neutroni come entità indipendenti, e riflette perciò la concezione di «protone neutro» a cui si rifà espressamente Majorana.

Limitando lo scambio alle sole posizioni, Majorana conclude che la particella α è stabile. Il fisico italiano apporta un'ulteriore modificazione alla teoria di Heisenberg, grazie alla quale può evitare di introdurre artificiosamente una distanza minima tra le particelle per garantire che le forze si saturino: il nucleo avrà densità costante, indipendentemente dalla propria massa, e il volume e l'energia nucleari saranno semplicemente proporzionali al numero di particelle.

Nonostante contenesse correzioni alla sua teoria, dunque, il contributo di Majorana piacque molto a Heisenberg; Majorana lo riferisce per lettera al padre, ma l'apprezzamento trapela anche, e in maniera evidentissima, dagli atti del settimo Congresso Solvay: sorprende molto, infatti, constatare che il fisico tedesco, a cui si deve la prima versione data alle stampe, nel presentare la teoria delle forze di scambio ricorre *quasi sempre* a frasi del tipo «d'apres Majorana», «en suivant l'exemple de Majorana», «comme y a insisté Majorana», «nous choisirons avec Majorana»,[63] menzionando soltanto in rarissime occasioni e in maniera del tutto defilata il proprio personale contributo. È talmente insistente questo richiamo al fisico catanese che sembra quasi lasciar trapelare un debito di riconoscenza verso il collega che aveva apportato alcune correzioni alla sua teoria.

Che nell'ottobre del 1933 Heisenberg avesse definitivamente abbracciato la linea di pensiero di Majorana trova eco anche in una sua critica alla presenza di elettroni nel nucleo: «L'affirmation: les électrons figurent comme constituants nucléaires, ne possède aucune signification définie en dépit du fait rappelé plus haut que beaucoup de noyaux émettent des rayons β».[64]

Una volta applicata la meccanica quantistica al nucleo mediante forze di scambio, rimaneva da spiegare il decadimento β, ossia come potessero essere espulsi elettroni negativi da un nucleo composto di particelle cariche positivamente e di «protoni neutri». Majorana, per risolvere le difficoltà teoriche legate al nucleo e in particolare alla presenza di elettroni nucleari aveva scelto di «tentare di stabilire la legge di interazione tra le particelle elementari in base a *soli criteri di semplicità*».[65] Tra la fine del 1933 e il 1934 Fermi pubblicò la teoria che superava i problemi inerenti al decadimento β. Egli escluse la complessità del neutrone (e, diversamente dal modello discusso alcuni mesi prima con Wataghin, quella del protone). Anche Fermi si ispirò a criteri di semplicità: «La via *più semplice* – scrive il fisico romano – per la costruzione di una teoria che permetta una discussione quantitativa dei fenomeni in cui intervengono elettroni nucleari, sembra [...] doversi ricercare nell'ipotesi che gli elettroni non esistano come tali nel nucleo prima dell'emissione β».[66] Poi ancora: «Il formalismo matematico *più semplice* per costruire una teoria in cui il numero di particelle leggere (elettroni e neutrini) non sia necessariamente costante, si ha nel metodo di Dirac-Jordan-Klein delle "ampiezze di probabilità quantizzate"»; e di nuovo, la scelta da adottare per l'energia di interazione è «la *più semplice*».[67]

Fermi ammette che il nucleo non contenga né elettroni né neutrini, perché diversamente non si potrebbe giustificare facilmente come particelle così leggere possano essere trattenute entro dimensioni dell'ordine di quelle nucleari: ritiene «più appropriato» considerare che i costituenti fondamentali siano le sole particelle pesanti, protoni e neutroni. Tale circostanza costringe però ad affrontare la questione dell'origine degli elettroni emessi nel decadimento β: nella teoria di Fermi, il neutrone si trasforma in un protone, e un elettrone e un 'neutrino' sono creati *ex novo*.

Grazie alla teoria pubblicata da Fermi, che si va ad aggiungere a quella di Heisenberg e Majorana, è dunque possibile non solamente applicare la meccanica quantistica al nucleo mediante forze di scambio, ma anche descrivere il fenomeno della radioattività β con la completa esclusione degli elettroni dal nucleo. In questo faticoso percorso verso la concezione del neutrone come protone neutro è stato fondamentale l'apporto di due fisici dell'Istituto di via Panisperna: quello di Majorana, prima, poi quello di Fermi.

Amaldi sostiene che, da quando Gentile nel 1928 si era occupato della teoria dei satelliti di Rutherford, «l'idea che vi potessero essere in natura dei corpuscoli neutri di dimensioni subatomiche era per così dire rimasta nell'aria anche a Roma».[68] Sulla base della breve rassegna presentata in questa nota sui principali contributi alla questione della costituzione del neutrone, possiamo dire qualcosa in più: il Regio Istituto Fisico dell'Università di Roma era stato nei primi anni Trenta baluardo e centro di irradiazione che aveva difeso e promosso nel panorama scientifico internazionale, fino alla definitiva accettazione, l'idea secondo la quale il neutrone non andava inteso più nella concezione di Harkins, Rutherford e della scuola di Cambridge, ossia come combinazione di protone ed elettrone, bensì nella nuova accezione di «protone neutro», ossia particella indipendente, o fondamentale.

*SUMMARY*

*In this paper the coming into light of the neutron is discussed.*

*A very brief review is reported at first of the experimental steps that revealed the existence of a particle of mass 1 and zero charge. Joliot-Curies' idea of considering the penetrating radiation from beryllium as to consist of high energy gamma rays was not so ingenuous as it is often depicted: previous experimental results had already pointed to an electromagnetic component; in some respect, a quantitative agreement also existed. The Joliot-Curies, moreover, relied on the many Compton electron tracks in the expansion chamber.*

*The neutron was foreseen in 1920 both by Rutherford and, still a bit earlier than him, by Harkins. Both of them thought of such a particle as to consist of one proton and one electron in strict combination – though they did not at first give any name to it. The circumstance is stressed that, when the neutron of Harkins and Rutherford was finally observed in 1932 by Chadwick, there was still room for considering it as a compound particle. So, Chadwick's experiments did not suffice to settle the problem of the constitution of nuclei. In particular, they did not settle at all the problem concerning the presence of electrons in the nucleus.*

*That question, whether considering the neutron as a fundamental particle or considering it as a compound particle, still bothered Chadwick himself for more than one year. He was not*

*the only one who wondered about the presence of electrons in the nuclei, and then inside the neutron. Even Heisenberg, as is well known, was ambiguous about the presence of electrons in the nucleus, in his first papers about exchange interactions – where he refers to null-spin 'electrons', obeying Bose statistics.*

*Majorana – who, as soon as he had heard about Joliot-Curies' experiments, guessed the existence of a 'neutral proton' – in 1933 published a paper containing some 'corrections' to Heisenberg's theory. Heisenberg deeply appreciated Majorana's work. A very simple analysis highlights that Heisenberg's original version of exchange interactions clearly reflected the view that the nucleus contained some sort of 'electrons', while Majorana's version that the nucleus was free from electrons.*

*Nevertheless, if the nucleus were to be free from electrons, still, the emission of electrons from the nucleus itself needed to be accounted for in some way. Fermi's theory of the **b**-decay – showing in its procedures some 'philosophical' similarity with Majorana's ones – settled the question. Thanks to exchange interactions and **b**-decay theory, quantum mechanics could now be applied to the nucleus, and – which is the same – the nucleus could be considered free from electrons; so could the neutron, which was allowed to be treated as a fundamental particle. That is, the neutron came into light in Cambridge, but its formal birth certificate as an independent particle was provided in Rome.*

---

[1] J. Chadwick, *Possible Existence of a Neutron*, «Nature», CXXIX, 1932, p. 312.

[2] W.W. Bothe e H. Becker, *Eine Kern-**g**-Strahlung bei leichten Elementen*, «Naturwissenschaften», XVII, 1930, p. 705; W.W. Bothe e H. Becker, *Künstliche Erregung von Kern-**g**-Strahlen*, «Zeitschrift für Physik», LXVI, 1930, pp. 289-306.

[3] W.W. Bothe e H. Fränz, *Atomtrümmer, reflektierte **a**-Teilchen und durch **a**-Strahlen erregte Röntgenstrahlen*, «Zeitschrift für Physik», IL, 1928, pp. 1-26.

[4] Cfr. J. Chadwick, J.E.R: Constable e E.C. Pollard *Artificial Disintegration by **a**-Particles*, «Proceedings of the Royal Society», CXXX, 1931, pp. 463-89, qui pp. 480-81. Il difetto di massa consiste nella differenza tra la massa del nucleo e quella dei suoi componenti liberi; esprime l'energia con cui sono tenuti uniti i vari costituenti del nudeo, misurata in unità di massa secondo l'equivalenza relativistica tra massa ed energia. Il caso in questione, perciò, riguardava considerazioni sull'energia necessaria a tenere insieme le particelle che si riteneva costituissero i nuclei.

[5] B. Rossi, *Il problema della radiazione penetrante*, in *Convegno di fisica nucleare (ottobre 1931)*, Roma, Reale Accademia d'Italia, 1932, pp. 51-64.

[6] W.W. Bothe e W. Kolhörster, *Das Wesen der Höhenstrahlung*, «Zeitschrift für Physik», LVI, 1929, pp. 751-77.

[7] Tale sezione d'urto misura la probabilità d'interazione per effetto Compton di un fotone con un elettrone, ed è inversamente proporzionale al cammino medio percorso da un fotone prima che interagisca con un elettrone.

[8] W.W. Bothe, ***a**-Strahlen, Künstliche Kernumwandlung und -Anregung, Isotope*, in *Convegno di fisica nucleare (ottobre 1931)*, Roma, Reale Accademia d'Italia, 1932, cit. in nota 5, pp. 83-106; W.W. Bothe, *Bemerkungen Über die Ultra-Korpuskularstrahlung*, *ibid.*, pp. 153-54.

[9] I. Curie, *Sur le rayonnement **g** nucléaire excité dans le glucinium et dans le lithium par les rayons **a** du polonium*, «Comptes Rendus de l'Académie des Sciences», **193**, 1931, pp. 1412-14; F. Joliot, *Sur l'excitation des rayons **g** nucléaires du bore par les particules **a**. Énergie quantique du rayonnement **g** du polonium*, «Comptes Rendus de l'Académie des Sciences», **193**, 1931, pp. 1415-17.

[10] I. Curie, *Sur le rayonnement **g** nucléaire excité dans le glucinium et dans le lithium par les rayons **a** du polonium*, cit in nota 9, p. 1413. Il valore ottenuto da Irène Curie per il «coefficient d'absorption massique» nel piombo era pari a 0,013 cm$^2$/g, a fronte del precedente 0,02. Tale coefficiente era dato dal rapporto $\mu/\rho$ tra la costante di assorbimento $\mu$ divisa per la densità $\rho$ dell'assorbitore (dove la costante $\mu$ compariva nell'espressione $I(x)/I_0 = \exp(-\mu x)$, che forniva l'intensità della radiazione assorbita dopo un percorso x). Nel piombo, per esempio, un coefficiente di assorbimento «massico» pari a 0,1 cm$^2$/g corrispondeva a una radiazione la cui intensità si riduceva a metà in 0,61 cm di spessore, un coefficiente 0,01 a una riduzione a metà in 6,1 cm.

[11] I. Curie e F. Joliot, *Émission de protons de grande vitesse par les substances hydrogénées sous l'influence des rayons **g** trés pénétrants*, «Comptes Rendus de l'Académie des Sciences», **194**, 1932, pp. 273-75. I protoni emessi avevano energia di 4,5 MeV per il berillio e 2 MeV per il boro.

*satelliti di Rutherford*, cit. in nota 23, p. 348). Un primo passo verso l'uso di metodi quantistici per descrivere il nucleo e le particelle pesanti in esso contenute fu mosso, in quello stesso anno, da Gamow e da Gurney e Condon nel formulare la teoria del decadimento α (G. Gamow, *Zur Quantentheorie des Atomkernes*, «Zeitschrift für Physik», LI, 1928, pp. 204-12; R.W. Gurney e E.U. Condon, *Quantum Mechanics and Radioactive Disintegration*, «Physical Review», XXXIII, 1928, pp. 127-40).

[62] Cfr. *Structure et propriétés des noyaux atomiques*, cit. in nota 44, p. 300.

[63] «Secondo Majorana», «seguendo l'esempio di Majorana», «come ha insistito Majorana», «adotteremo, allo stesso modo di Majorana, … »; W. Heisenberg, *Considération théoriques générales sur la structure du noyau*, in *Structure et propriétés des noyaux atomiques*, cit. in nota 44, pp. 289-323, *passim*.

[64] «L'affermazione che gli elettroni compaiono come costituenti del nucleo non possiede alcun significato definito in aggiunta al fatto, ricordato prima, che molti nuclei emettono raggi β»; cfr. *Structure et propriétés des noyaux atomiques*, cit. in nota 44, pp. 292-93.

[65] E. Majorana, *Sulla teoria dei nuclei*, cit. in nota 60, p. 559 (riportato in E. Amaldi, *La vita e l'opera di Ettore Majorana (1906-1938)*, cit. in nota 25, p. 60). Questo corsivo e i tre che seguono sono aggiunti.

[66] E. Fermi, *Tentativo di una teoria dell'emissione dei raggi **b***, «La Ricerca Scientifica», anno IV, vol. II, 1933, pp. 491-95, *passim*.

[67] *Loc. cit.*

[68] E. Amaldi, *La vita e l'opera di Ettore Majorana (1906-1938)*, cit. in nota 25, p. XXII.